# Carbon nanoparticles in the acoustic field in the vicinity of the arc discharge

M.N. Shneider
*Department of Mechanical and Aerospace Engineering, Princeton University, Princeton, NJ 08544, USA*

The paper considers an effect of intensive ultrasound on the suspension of soot microparticles and nanoparticles in the inert gas, resulting in the coagulation of relatively large soot particles and leading to the improvement of the efficiency of production of nanoparticles, as has been observed in experiments. The effect of the particles charge on the possibility of coagulation is analyzed.

## I. Introduction

The arc discharge between the graphite electrodes burning in the atmosphere of inert gases is one of the standard methods of nanoparticle synthesis [1-5]. Typically, in theoretical models, it is assumed (see., e.g. [2,3,5]) that the synthesis starts in the areas where the density of carbon atoms and the catalyst is still high, and the temperature of a buffer gas is reduced down to 1000-2000 K.

The arc is a very powerful light emission source, which is close to the blackbody. This radiation is scattered and partly absorbed by the nanoparticles. In the areas where a synthesis can occur theoretically, the radiation intensity of the arc is still quite high, and therefore, one should expect that the temperature of nanoparticles that absorb radiation may considerably exceed the local ambient gas temperature. Radiative heating of nanoparticles in the radiation field of the arc has been considered in the recent paper [6].

In the peripheral region of arc burning in a high pressure inert gas, together with nanoparticles, a large number of microscopic soot particles is produced. Intensive soot generation significantly reduces the efficiency of the arc, as the technological process of production of fullerenes and other nanoparticles. Experimental studies have shown that exposure of intense ultrasound on the peripheral region of the arc leads to a noticeable increase in the efficiency of the synthesis of nanoparticles and to the reduction in the yield of soot (see., e.g. [7]). It is shown in this paper that the effect of ultrasound on the suspension of soot microparticles and nanoparticles in the inert gas results in the coagulation of soot particles, practically without affecting the nanoparticles. This effect contributes to the improvement of the efficiency of the nanoparticles generation, as has been observed in experiments [7].

## II. Effect of Ultrasound on the Mixture of Nanoparticles and Soot Microparticles

### A. Neutral particles

There are experimental works in which the effect of ultrasound on the yield of fullerenes and nanoparticles formed in the vicinity of the arc with carbon electrodes burning in an inert gas, was studied, showing a simultaneous decrease in the output of soot particles (see., e.g., [7]). These results seem surprising, since acoustic effects on nanoparticles with a size much smaller than the mean free path of the atoms of a buffer gas in which ultrasound is maintained are not expected. However, in our opinion, these results have a simple explanation.

It is known that in the ultrasonic field acting on the gas with suspended microparticles it is possible for the particles to coagulate, i.e. form larger particles from smaller ones. The interaction of spherical particles in a gas in a sound wave is determined by the theory of Köenig [8]. The radial component of the Köenig force have the form [9]:

$$F_r = \frac{3}{4}\frac{\pi\rho a_1^3 a_2^3 u_0^2}{L^4}\left(3\cos 2\theta + 1\right). \quad (1)$$

Here $u_0$ is the amplitude of the the oscillatory velocity in the sound wave; $\rho$ is the density of the medium; $a_1, a_2$ are the radii of the interacting spheres; $L$ is the distance between the centers of the particles; $\theta$ is the angle between the center line and the wave vector of the acoustic wave. The amplitude of the oscillatory velocity in the sound wave is related to the ultrasound intensity $I_s$, its angular frequency, $\omega$, and amplitude of the oscillations $A_0$ :

$$I_s = \tfrac{1}{2}\rho c_s \omega^2 A_0^2 = \tfrac{1}{2}\rho c_s u_0^2 , \tag{2}$$

where $c_s$ is the speed of sound in the gas.

Depending on the angle $\theta$, the particles can repel or attract each other. The maximum force of attraction (sometimes called the Bernoulli force) occurs when the line connecting the centers of the particles is perpendicular to the direction of propagation of the sound wave, that is, at $\theta = \pi / 2$ :

$$F_{r,m} = -\frac{3}{2}\frac{\pi \rho a_1^3 a_2^3 u_0^2}{L^4} . \tag{3}$$

The Koenig theory is valid for relatively large distances between the interacting particles, $L >> a_1 + a_2$. For smaller distances between the particles the perturbations of higher order should be taken into account. The corresponding correction has been received by Bjerknes [10,9]

$$\Delta F_r = -\frac{3}{4}\frac{\pi \rho a_1^3 a_2^3 (a_1^3 + a_2^3) u_0^2}{L^7}\left(3\cos^2\theta + 1\right). \tag{4}$$

If the particles are charged, the electrostatic interaction between them with the Coulomb force should be taken into account:

$$F_q = \frac{1}{4\pi\varepsilon_0}\frac{q_1 q_2}{L^2} . \tag{5}$$

In general, the equation of motion of spherical particles along a line connecting their centers is given by

$$m_{p_{1,2}} \frac{d^2 L}{dt^2} = (F_q + F_r + \Delta F_r) + F_{s_{1,2}} . \tag{6}$$

In a sufficiently dense gas when the particle size greatly exceeds the mean free path of the gas molecules, the retarding force $F_{s_{1,2}}$ acting on spherical particles is determined by the Stokes formula

$$F_{s_{1,2}} = -6\pi\eta a_{1,2} \frac{dL}{dt} , \tag{7}$$

where $\eta$ is the dynamic viscosity of the buffer gas. In this case, the particles are moving with the instantaneous drift velocity, which is determined by the balance of forces acting on the particle:

$$\frac{dL}{dt} = (F_q + F_r + \Delta F_r) / 6\pi\eta a . \tag{8}$$

The drift approximation is quite applicable to the sub-micron and micron-sized particles in helium at atmospheric pressure and relatively low temperatures < 3000 K, as, for example, at $p = 760$ Torr and $T = 298$ K, the mean free path $l_m \approx 0.2\ \mu$m .

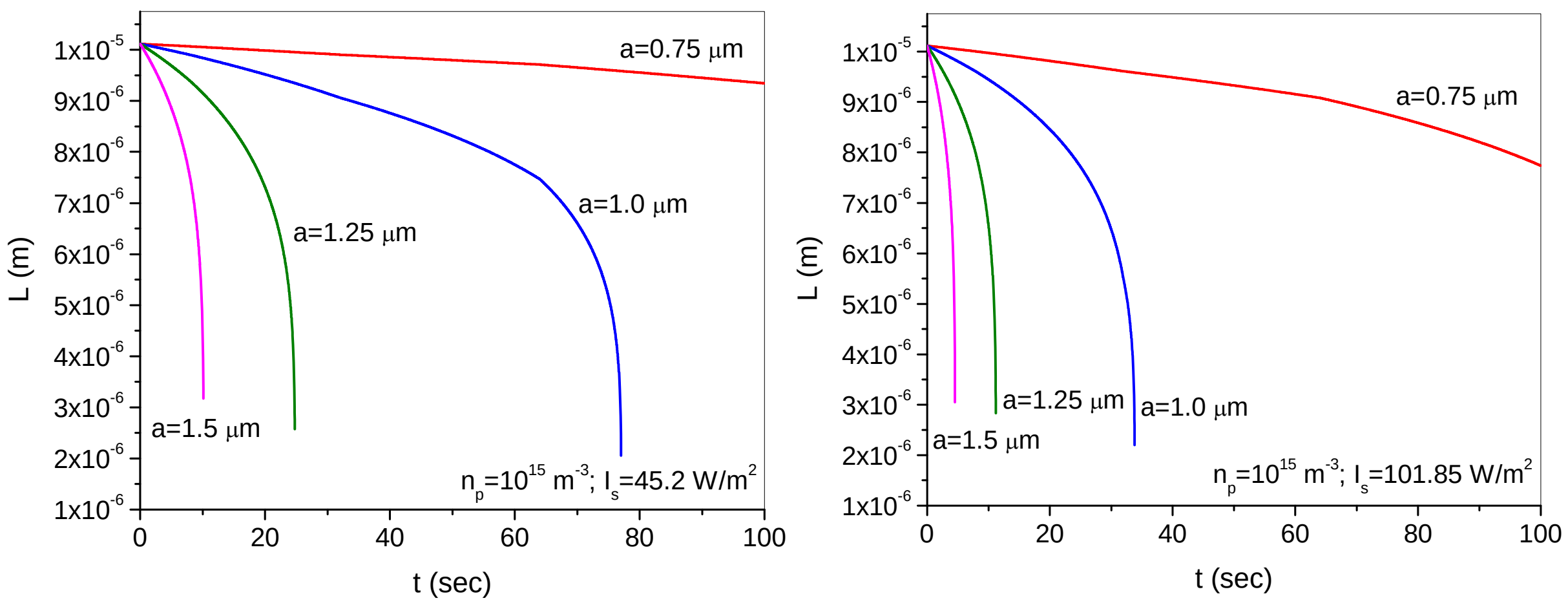

FIG. 1. Distance between identical neutral microscopic soot particles in helium at $p = 760$ torr, $T = 1000$ K at $\theta = \pi/2$ for various sizes of particles and ultrasound at intensities (a) $I_s = 45.2$ W/m² ($u_0 = 1$ m/s) and (b) $I_s = 101.85$ W/m² ($u_0 = 1.5$ m/s).

Examples of calculations for equal neutral microscopic soot particles in helium at $p = 760$ Torr, $T = 1000$K for various particle sizes and intensities of ultrasound at $\theta = \pi/2$ are shown in Fig. 1. It is assumed for definiteness that the initial concentration of particles is $n_p = 10^{15}$ m$^{-3}$. The dynamic viscosity of helium $\eta(T)$ was taken from [11].

When approaching to a distance $L \approx a_1 + a_2$, the particles stick together under the influence of short-range van der Waals forces. For larger particles, the effect is stronger, and they will be brought together in a few seconds or even less. Thereafter, they fall out of the volume under the influence of gravity. (So, by the way, runs a standard method of ultrasonic cleaning of gases.) With increasing of the volume of the particles, formed due to the coagulation, their loss is accelerated. Since, when the gravity force $Mg = \frac{4}{3}\pi a^3 \rho_p g$ is balanced by the frictional force (7), the corresponding drift velocity $u_g = \frac{2\rho_p a^2 g}{9\eta} \propto a^2$, where $\rho_p$ is the density of the particles.

Thus, more fullerenes and nanoparticles remain in a volume to which ultrasound practically has no effect. These fullerenes have less chance to disappear by sticking to larger particles, as a result of Brownian motion, because of a significant decrease in the concentration of soot particles in the volume.

Note that we have considered the dynamics of interacting particles in a traveling ultrasonic wave. In a standing sound wave the suspended particles are affected by the King force [12], that moves particles to the antinodes of the standing wave. It also leads to the coagulation and precipitation of particles. As the King force in the standing wave $\propto a^3$, it is also selective. Thus, larger particles are coagulated quicker and the coagulation process almost does not affect the particles of the size much smaller than the mean free path of gas molecules. In this brief paper we will not consider the processes in the standing wave in more details.

**B. Charged particles**

We have considered the case of neutral particles. In the plasma, or as a result of thermionic emission in the radiation field of the arc, particles are charged. In plasma, usually the particles acquire a negative charge, charging to the local floating potential. However, as a result of the thermionic emission, the particle loses electrons, i.e. it becomes positively charged. The Coulomb repulsion of the particles may be substantially greater than the Köenig force and the agglomeration of the microparticles occurs at increased ultrasound intensities or becomes completely impossible. Incidentally, charging of the agglomerates as a result of the thermionic emission caused by laser heating can lead to their disintegration in Laser Induced Incandescense (LII) experiments [13].

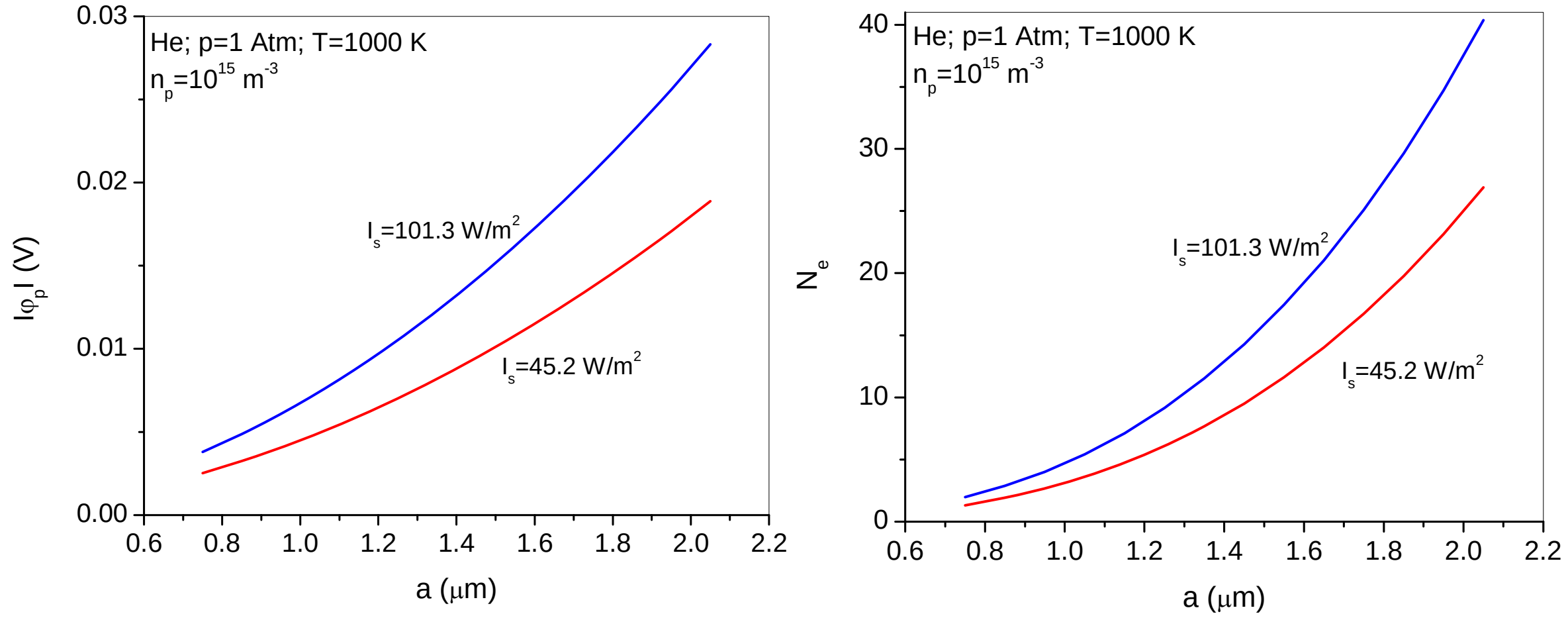

FIG. 2. Critical values of the potential (a) and charge (b) for the coagulation of identical charged soot particles of different sizes in helium at $p = 760$ torr, $T = 1000$ K at $\theta = \pi/2$ and ultrasound intensities $I_s = 45.2$ W/m$^2$ and $I_s = 101.85$ W/m$^2$

For simplicity, consider the background plasma of relatively low density, where the Debye length is greater than the average distance between the soot particles. For each ultrasound intensity and concentration of micro-particles of a certain size, there is a critical value of their charge (electrostatic potential), which still allows coagulation in an acoustic field. This limiting value of the potential can be determined when the absolute values of the Köenig attractive force (3) and Coulomb repulsive force (5) are equal. Charged particles approach each other by the action of the acoustic field, when

$$F_q \le |F_r| . \tag{9}$$

For identical particles, regardless of the sign of the charge,

$$|q| \le |q_c| = \frac{\pi a^3 u_0}{L}(6\varepsilon_0 \rho)^{1/2} , \; N_e = |q_c|/e \tag{10}$$

where $N_e$ is the corresponding number of electrons to be acquired (or lost) by the particle. Or, from the relation between the charge of the particle, its capacity and the potential $\varphi = q/C$, where $C = 4\pi\varepsilon_0 a$ is the capacity of the particle, the coagulation is possible at

$$|\varphi| \le |\varphi_c| = \frac{a^2 u_0}{2L}(3\rho / 2\varepsilon_0)^{1/2} . \tag{11}$$

The results of calculations with the estimated formulas (10),(11), presented in Fig.2, show that a very small charge acquired by the soot particles is sufficient to suppress the coagulation process. However, for the same conditions, with increasing ultrasound intensity, the coagulation again becomes possible.

## III. Conclusions

- Coagulation of soot particles in the ultrasonic field is possible, resulting in a decrease in the concentration of soot particles and increase in production efficiency of fullerenes and nanoparticles.
- The threshold for the ultrasound intensity required for the coagulation depends on charge acquired by the soot particles, particle sizes and background gas parameters

## Acknowledgements

I would like to thank Dr. Yevgeny Raitses, Dr. Igor Kaganovich, and Mr. James Mitrani for their interest in this work and fruitful discussions. This work was supported by the U.S. Department of Energy, Office of Science, Basic Energy Sciences, Materials Sciences and Engineering Division.